\documentclass[twocolumn,english,pra,aps,superscriptaddress]{revtex4-1}
\usepackage[T1]{fontenc}
\usepackage[latin9]{inputenc}
\usepackage{amsmath}
\usepackage{graphicx}

\makeatletter
\usepackage{babel}

\makeatother

\usepackage{babel}
\begin{document}

\title{The Bose-Fermi duality in quantum Otto heat engine with trapped repulsive
Bosons}

\date{\today}

\author{Jinfu Chen}

\address{Beijing Computational Science Research Center, Beijing 100193, China}

\address{Graduate School of China Academy of Engineering Physics, No. 10 Xibeiwang
East Road, Haidian District, Beijing, 100193, China}

\author{Hui Dong}
\email{hdong@gscaep.ac.cn}

\selectlanguage{english}%

\address{Graduate School of China Academy of Engineering Physics, No. 10 Xibeiwang
East Road, Haidian District, Beijing, 100193, China}

\author{Chang-Pu Sun}
\email{cpsun@csrc.ac.cn}

\selectlanguage{english}%

\address{Beijing Computational Science Research Center, Beijing 100193, China}

\address{Graduate School of China Academy of Engineering Physics, No. 10 Xibeiwang
East Road, Haidian District, Beijing, 100193, China}
\begin{abstract}
Quantum heat engine with ideal gas has been well studied, yet the
role of interaction was seldom explored. We construct a quantum Otto
heat engine with $N$ repulsive Bosonic particles in a 1D hard wall
box. With the advantage of exact solution using Bethe Ansatz, we obtain
not only the exact numerical result of efficiency in all interacting
strength $c$, but also analytical results for strong interaction.
We find the efficiency $\eta$ recovers to the one of non-interacting
case $\eta_{\mathrm{non}}=1-(\text{\ensuremath{L_{1}/L_{2}}}){}^{2}$
for strong interaction with asymptotic behavior $\eta\sim\eta_{\mathrm{non}}-4(N-1)L_{1}\left(L_{2}-L_{1}\right)/(cL_{2}^{3})$.
Here, $L_{1}$ and $L_{2}$ are two trap sizes during the cycle. Such
recovery reflects the duality between 1D strongly repulsive Bosons
and free Fermions. We observe and explain the appearance of a minimum
efficiency at a particular interacting strength $c$, and study its
dependence on the temperature.
\end{abstract}
\maketitle

\section{Introduction}

In classical thermodynamics, piston model with ideal gas serves as
a prototype to realize heat engines with the different cycles, such
as Carnot and Otto cycle \cite{book:18204,book:18254}. The non-interacting
gas makes it feasible to obtain very simple results for efficiency
as well as other properties \cite{book:18204}. Such simplicity also
enables direct extensions of similar discussions in quantum region
to show unique features of quantum thermodynamics with single particle
as well as few identical particles without interaction \cite{Quan2007,Dong2011,Kim2011,Cai2012,Gelbwaser-Klimovsky2015,Quan2009}
or with interaction only changing the energy for the ground state
\cite{Jaramillo2016,Deng2018}. The difficulty arises when the interaction
changes the energy spectrum  for a quantum heat engine. Recently,
Bengtsson et.al explores the effect of the attractive interaction
with exact numerical simulation and show the increase of work output
in the Szilard engine \cite{Bengtsson2018}. However, it remains unclear
how the work conversion is affected by interactions in the widely
used heat engine cycles, e.g., quantum Otto cycle. 

In order to show the effect of interaction on the efficiency, we construct
a quantum Otto heat engine with 1D repulsive Bose gas in a hard wall
box \cite{Gaudin1971,Batchelor2005,Oelkers2006}. Quantum Otto cycle
is a simple and feasible cycle in quantum thermodynamics \cite{Quan2007,Newman2017},
and has been studied concretely in many quantum systems \cite{Thomas2011,Huebner2014,Altintas2014,Zhang2014a,Ivanchenko2015,Jaramillo2016,Beau2016,Deng2018,Huang2018}.
The advantage of our model is its exact solution with Bethe\textquoteright s
ansatz \cite{Yang1967,Gaudin1971}, which allows the analytical results
to show the effect of interaction in quantum thermodynamics. We find
that the efficiency of the heat engine first decreases, reaches to
the minimum value and finally recovers to the initial value along
with the increasing of the interacting strength $c$. For strong interacting
strength, the recovery of the efficiency is explained by the Bose-Fermi
duality \cite{Girardeau1960,Wang2018}. We observe a dip for the efficiency
with particular interacting strength. We also study how temperature
affects the dip of the efficiency. 

This paper is organized as follows. In Sec. II, we introduce the solution
of 1D interacting Bose gas in a hard wall box by Bethe ansatz, and
build the quantum Otto heat engine on this model. In Sec. III, we
give the asymptotic efficiency for large interacting strength and
the numerical efficiency for any interacting strength. The recovery
of the efficiency for large interacting strength reflects the Bose-Fermi
duality between 1D strongly repulsive Bosons and free Fermions. We
associate the efficiency with the ratio for different state, and study
the efficiency for different temperature.

\section{Otto engine with repulsive Bosons}

In this section, we design a quantum Otto heat engine with 1D repulsive
Bose gas in a hard wall box. The efficiency of the quantum Otto cycle
for different interacting strength $c$ is studied to explore the
effect of the interaction on the quantum heat engine.

The Hamiltonian for $N$ repulsive Bosonic particles in a 1D hard
wall box is 
\begin{equation}
H(L,c)=\sum_{i=1}^{N}\frac{p_{i}^{2}}{2m}+\frac{c}{m}\sum_{i<j}\delta\left(x_{i}-x_{j}\right)+V\left(\{x_{i}\}\right),\label{eq:1}
\end{equation}
where $p_{i}$ and $x_{i}$ are the momentum and coordinate for $i$-th
particle with mass $m$. The interacting strength $c$ is positive
for repulsive interaction. The trap potential $V\left(\{x_{i}\}\right)$
is infinite square potential
\begin{equation}
V\left(\{x_{i}\}\right)=\begin{cases}
0 & \forall\,0\leq x_{i}\leq L\\
\infty & \exists\,x_{i}<0,\:\mathrm{or\:}x_{i}>L.
\end{cases}\label{eq:2}
\end{equation}
Eigenstates can be obtained with Bethe Ansatz \cite{Gaudin1971} for
the current trap. The eigenstate is written as\textbf{ $\psi_{\{k_{i}\}}(\{x_{i}\})=\sum_{P}a(P)\exp(i\sum_{l=1}^{N}k_{P(l)}x_{l}),\,0\leq x_{1}\leq x_{2},...\leq x_{N}\leq L$}
, with the superposition coefficient $a(P)$. We are interested in
the thermodynamic property and skip the concrete form of $a(P)$,
whose explicit form can be found in Ref. \cite{Gaudin1971}. The boundary
condition gives the self-consistent equation for the wave vectors
as

\begin{equation}
k_{i}L=\pi n_{i}+\sum_{j\ne i}(\arctan\frac{c}{k_{i}-k_{j}}+\arctan\frac{c}{k_{i}+k_{j}}).\label{eq:3}
\end{equation}
The eigenstate $\left|\{n_{i}\}\right\rangle $ is represented by
a set of ordered number \textbf{$1\leq n_{1}\leq n_{2}\leq...\leq n_{N}$,
}and the wave vectors satisfy\textbf{ $1<k_{1}<k_{2}<...<k_{N}$.
}The corresponding energy for the eigenstate $\left|\{n_{i}\}\right\rangle $
is 
\begin{equation}
E_{\{n_{i}\}}^{(L)}=\frac{1}{2m}\sum_{i=1}^{N}k_{i}^{2}.\label{eq:energy}
\end{equation}

With the certain trap size $L$ and interacting strength $c$, the
density matrix for the system at the equilibrium state with temperature
$T$ is $\rho=\sum_{\{n_{i}\}}p_{\{n_{i}\}}(T,L,c)\left|\{n_{i}\}\right\rangle \left\langle \{n_{i}\}\right|$.
The probability $p_{\{n_{i}\}}(T,L,c)$ on the eigenstate $\left|\{n_{i}\}\right\rangle $
is
\begin{equation}
p_{\{n_{i}\}}(T,L,c)=\frac{e^{-\frac{E_{\{n_{i}\}}^{(L)}}{k_{B}T}}}{Z(T,L,c)},\label{eq:4-prob}
\end{equation}
with the partition function $Z(T,L,c)=\sum_{\{n_{i}\}}\exp[-E_{\{n_{i}\}}^{(L)}/kT]$.
The internal energy of the system is $U(T,L,c)=\sum_{\{n_{i}\}}p_{\{n_{i}\}}E_{\{n_{i}\}}^{(L)}.$ 

\begin{figure}[h]
\includegraphics[scale=0.75]{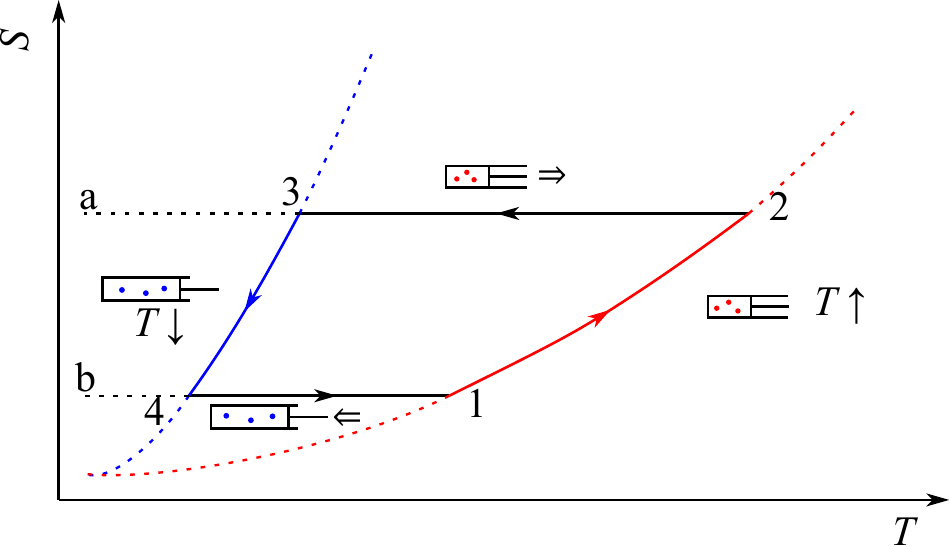}\caption{(Color online) Entropy-temperature ($S-T$) diagram for quantum Otto
heat engine. The red and blue solid lines are isochoric processes
contacting to the cold and hot reservoir with the corresponding trap
size as $L_{1}$ and $L_{2}$, while the black solid lines are adiabatic
processes.}

\label{fig:1_ST_diagram} 
\end{figure}

The Otto cycle consists four strokes similar to the single particle
Otto cycle \cite{Quan2007}, illustrated on $S-T$ diagram in Figure
\ref{fig:1_ST_diagram}. Here, the entropy is evaluated with the von
Neumann entropy $S=-\mathrm{Tr}[\rho\ln\rho]$ for all later discussions.
The four strokes are specified as follows, 
\begin{itemize}
\item Stroke I ($1\rightarrow2$) : Isochoric heating. Initially, the system
does not necessarily stay at equilibrium thermal state. The internal
energy of the system at state 1 is $U_{1}=\sum_{\{n_{i}\}}p_{\{n_{i}\}}^{(1)}E_{\{n_{i}\}}^{(L_{1})}$
. With the fixed trap size $L_{1}$, the temperature increases slowly
enough to allow the system in thermal equilibrium state with temperature
$T_{2}$ . The internal energy at state 2 is $U_{2}=\sum_{\{n_{i}\}}p_{\{n_{i}\}}^{(2)}E_{\{n_{i}\}}^{(L_{1})}$
with the equilibrium occupation $p_{\{n_{i}\}}^{(2)}=p_{\{n_{i}\}}(T_{2},L_{1},c)$.
The internal energy increases by absorbing heat from the hot reservoir
$Q_{1}=U_{2}-U_{1}>0$. 
\item Stroke II ($2\rightarrow3$): Quantum adiabatic expansion. During
the process, the system is isolated from any reservoir, and the trap
size increases from $L_{1}$ to $L_{2}$ slowly in order to keep the
occupation number unchanged, namely $p_{\{n_{i}\}}^{(3)}=p_{\{n_{i}\}}^{(2)}$.
In this process, the internal energy decreases from $U_{2}=\sum_{\{n_{i}\}}p_{\{n_{i}\}}^{(2)}E_{\{n_{i}\}}^{(L_{1})}$
to $U_{3}=\sum_{\{n_{i}\}}p_{\{n_{i}\}}^{(3)}E_{\{n_{i}\}}^{(L_{2})}$
to export work $W_{1}=U_{3}-U_{2}=\sum_{\{n_{i}\}}p_{\{n_{i}\}}^{(2)}(E_{\{n_{i}\}}^{(L_{2})}-E_{\{n_{i}\}}^{(L_{1})})<0$.
After the expansion, the system reaches generally a non-equilibrium
state, except that the energy levels shift homogeneously.
\item Stroke III ($3\rightarrow4$): Isochoric cooling. Similar to stroke
I, the trap size is fixed at $L_{2}$. The temperature decreases slowly
enough to allow the system in thermal equilibrium state with temperature
$T_{4}$. The occupation is $p_{\{n_{i}\}}^{(4)}=p_{\{n_{i}\}}(T_{4},L_{2},c)$,
and the internal energy is $U_{4}=\sum_{\{n_{i}\}}p_{\{n_{i}\}}^{(4)}E_{\{n_{i}\}}^{(L_{2})}$.
The internal energy decreases by releasing heat to the cold reservoir
$Q_{2}=U_{4}-U_{3}<0$. 
\item Stroke IV ($4\rightarrow1$): Quantum adiabatic compressing. Similar
to stroke II, the system is isolated from any reservoir, and the trap
size decreases from $L_{2}$ to $L_{1}$ slowly to keep the probability
$p_{\{n_{i}\}}$ as a constant, namely, $p_{\{n_{i}\}}^{(1)}=p_{\{n_{i}\}}^{(4)}$.
The system reaches a non-equilibrium state as the initial state of
stroke I. The internal energy increases by performed work $W_{2}=U_{1}-U_{4}=\sum_{\{n_{i}\}}p_{\{n_{i}\}}^{(4)}(E_{\{n_{i}\}}^{(L_{1})}-E_{\{n_{i}\}}^{(L_{2})})>0$.
\end{itemize}
The extracted work for the quantum Otto cycle $W_{out}=-W_{1}-W_{2}=Q_{1}-\left|Q_{2}\right|$
should be positive to ensure a valid heat engine instead of refrigerator.
For ideal gas, the positive work condition gives the requirement $T_{2}>\left(L_{2}/L_{1}\right)^{2}T_{4}$
\cite{Quan2007}. Such condition for repulsive Bosons is complicated
without a simple formula. Yet, in all later discussion, we carefully
choose the temperature $T_{2},T_{4}$ to allow the positive work. 

The efficiency $\eta=1-\left|Q_{2}\right|/Q_{1}$ is written explicitly
as

\begin{equation}
\eta=\frac{\sum_{\{n_{i}\}}\left(p_{\{n_{i}\}}^{(2)}-p_{\{n_{i}\}}^{(4)}\right)(E_{\{n_{i}\}}^{(L_{1})}-E_{\{n_{i}\}}^{(L_{2})})}{\sum_{\{n_{i}\}}\left(p_{\{n_{i}\}}^{(2)}-p_{\{n_{i}\}}^{(4)}\right)E_{\{n_{i}\}}^{(L_{1})}}.\label{eq:6}
\end{equation}
In the following numerical calculations, we use Eq. (\ref{eq:6})
to calculate the exact efficiency.

\section{The Efficiency and the Interaction}

For the large interacting strength $c$, we have the expansion of
Eq. (\ref{eq:3}) to the first order of $1/c$
\begin{equation}
k_{i}L=\pi(n_{i}+i-1)-\sum_{j\ne i}\left(\frac{k_{i}-k_{j}}{c}+\frac{k_{i}+k_{j}}{c}\right).
\end{equation}
The solution for the wave vector is $k_{i}=\pi\left(n_{i}+i-1\right)/\left(L+2\left(N-1\right)/c\right).$
Eq. (\ref{eq:energy}) gives the asymptotic energy for the eigenstate
$\left|\{n_{i}\}\right\rangle $ as

\begin{equation}
E_{\left\{ n_{i}\right\} }^{(L)}=\frac{\pi^{2}}{2m}\frac{\sum_{i=1}^{N}\left(n_{i}+i-1\right)^{2}}{\left(L+\frac{2\left(N-1\right)}{c}\right)^{2}}.\label{eq:1-1}
\end{equation}
 For the large interacting strength, the energy ratios for eigenstates
with different trap size have the same value
\begin{equation}
\frac{E_{\left\{ n_{i}\right\} }^{(L_{2})}}{E_{\left\{ n_{i}\right\} }^{(L_{1})}}=\left(\frac{L_{1}+\frac{2\left(N-1\right)}{c}}{L_{2}+\frac{2\left(N-1\right)}{c}}\right)^{2}.\label{eq:7ratio}
\end{equation}
Therefore, the internal energy for the initial state and the final
state of the quantum adiabatic processes has the same ratio as
\begin{equation}
\frac{U_{3}}{U_{2}}=\frac{U_{4}}{U_{1}}=\left(\frac{L_{1}+\frac{2\left(N-1\right)}{c}}{L_{2}+\frac{2\left(N-1\right)}{c}}\right)^{2}.
\end{equation}
 By Eq. (\ref{eq:6}), we obtain the efficiency as 
\begin{equation}
\eta=1-\left(\frac{L_{1}+\frac{2\left(N-1\right)}{c}}{L_{2}+\frac{2\left(N-1\right)}{c}}\right)^{2}.\label{eq:8}
\end{equation}
At the large interaction limit, we expand Eq. (\ref{eq:8}) to get
an asymptotic  efficiency to the first order of $1/c$
\begin{equation}
\eta\approx1-(\frac{L_{1}}{L_{2}})^{2}-\frac{4L_{1}\left(L_{2}-L_{1}\right)}{L_{2}^{3}}\frac{(N-1)}{c}.\label{eq:12}
\end{equation}
Such efficiency matches the one of non-interacting Fermions/Bosons,
which is the same for one-single particle quantum Otto heat engine
\cite{Quan2007}. The recovery the efficiency at strong coupling limit
to the ideal gas is caused by the duality between Fermions and Bosons
at 1D case \cite{Gaudin1971}. Such duality shows the match between
energy levels of strong repulsive interacting Bosons and non-interacting
Fermions, or verse vice. 

\begin{figure}[h]
\includegraphics[scale=0.2]{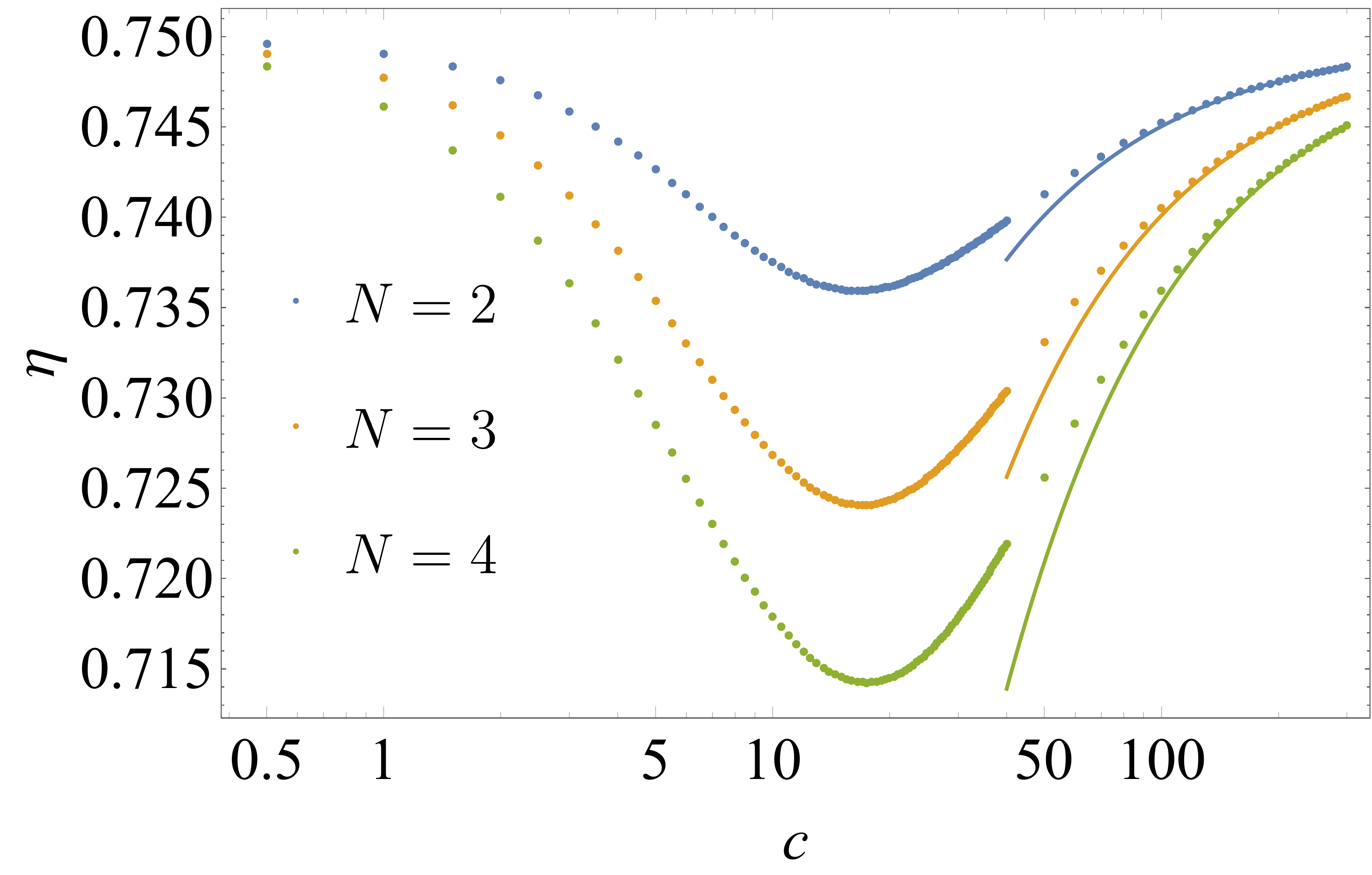}

\caption{(Color online) the log-linear plot for the efficiency $\eta$ for
the quantum Otto heat engine with different interacting strength $c$.
We consider three cases with the particle number as $N=2,\,3,\,4$,
and choose the temperature as $T_{2}=50,\,T_{4}=8.$ For all numerical
calculation, the mass and the cutoff of the quantum number are set
as $m=1$ and $n_{\mathrm{cut}}=20$ , and the trap size is always
set as $L_{1}=1,\:L_{2}=2.$ The solid line is the analytical result
of the asymptotic efficiency for large $c$ by Eq. (\ref{eq:1-1}),
while the dots are the exact numerical result.}
\label{2_loglinearplottwoparticles}
\end{figure}

To validate our result in Eq. (\ref{eq:8}, \ref{eq:12}), we compare
it to the exact numerical result in Fig. \ref{2_loglinearplottwoparticles}.
The efficiency for heat engine with different numbers $N=2,3,4$ of
Bosons are plotted as functions of interacting strength $c$. In the
simulation, we set the mass $m=1$, and choose a cutoff $n_{\mathrm{cut}}=20$
for the energy level index $n_{i}$, namely, $n_{i}\leq n_{\mathrm{cut}}$.
For the exact numerical calculation, we firstly calculate the energy
levels $E_{\left\{ n_{i}\right\} }^{(L_{1})}$ and $E_{\left\{ n_{i}\right\} }^{(L_{2})}$
by exactly solving Eq. (\ref{eq:3}) with the trap size as $L_{1}=1$
and $L_{2}=2$. Next, we calculate the probability $p_{\{n_{i}\}}^{(2)}$
and $p_{\{n_{i}\}}^{(4)}$ for the equilibrium state $2$ and $4$
with the temperature $T_{2}=50$ and $T_{4}=8$ for the hot and cold
reservoir respectively. And the exact efficiency is evaluated via
Eq. (\ref{eq:6}) with the probability $p_{\{n_{i}\}}^{(i)}$ and
the energy levels $E_{\left\{ n_{i}\right\} }^{(L_{i})}$.  In Fig
\ref{2_loglinearplottwoparticles}, we show that the numerical result
matches the analytical result by Eq. (\ref{eq:8}) well for the large
interacting strength, . 

\begin{figure}[h]
\includegraphics[scale=0.2]{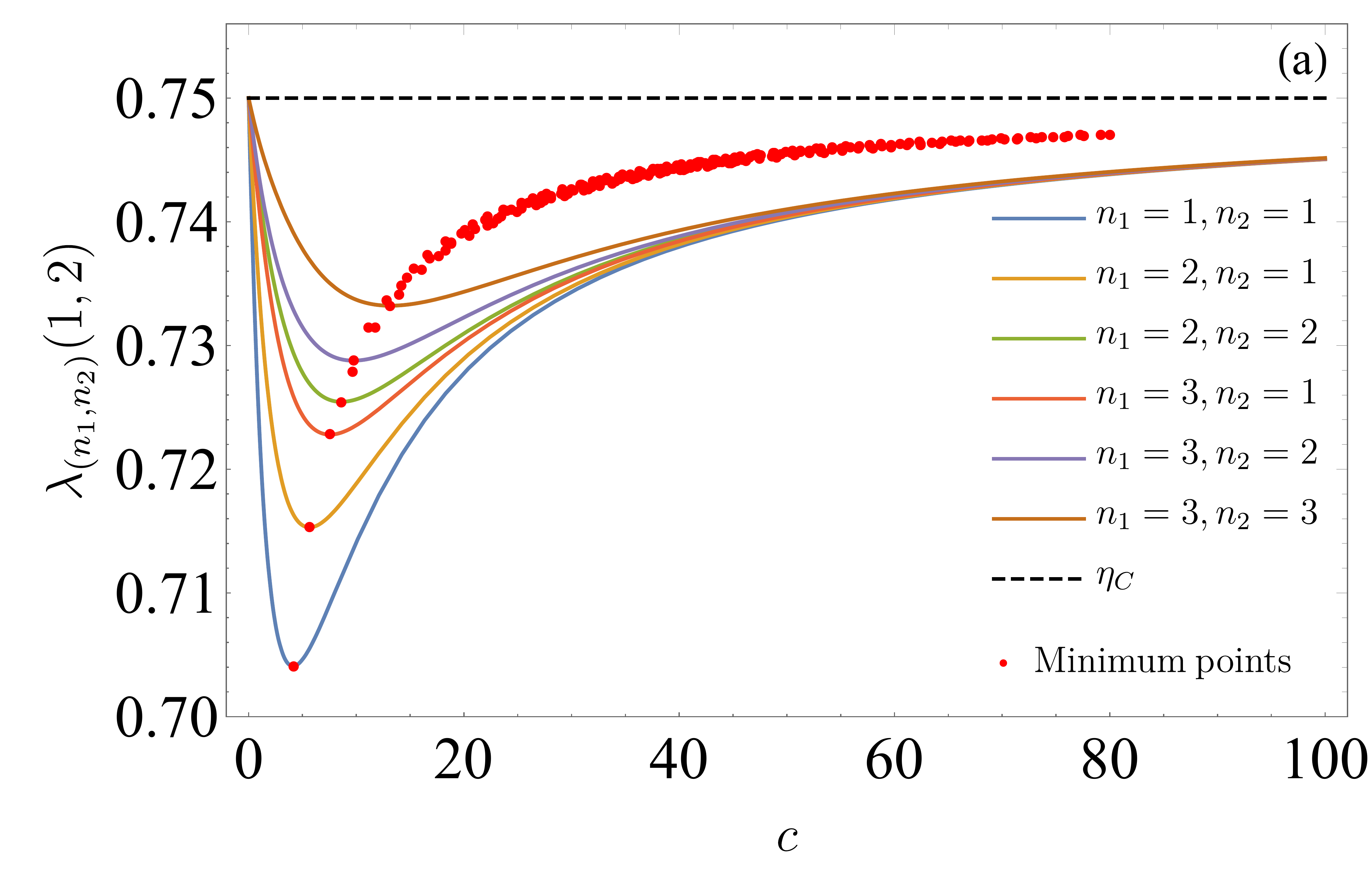}

\includegraphics[scale=0.2]{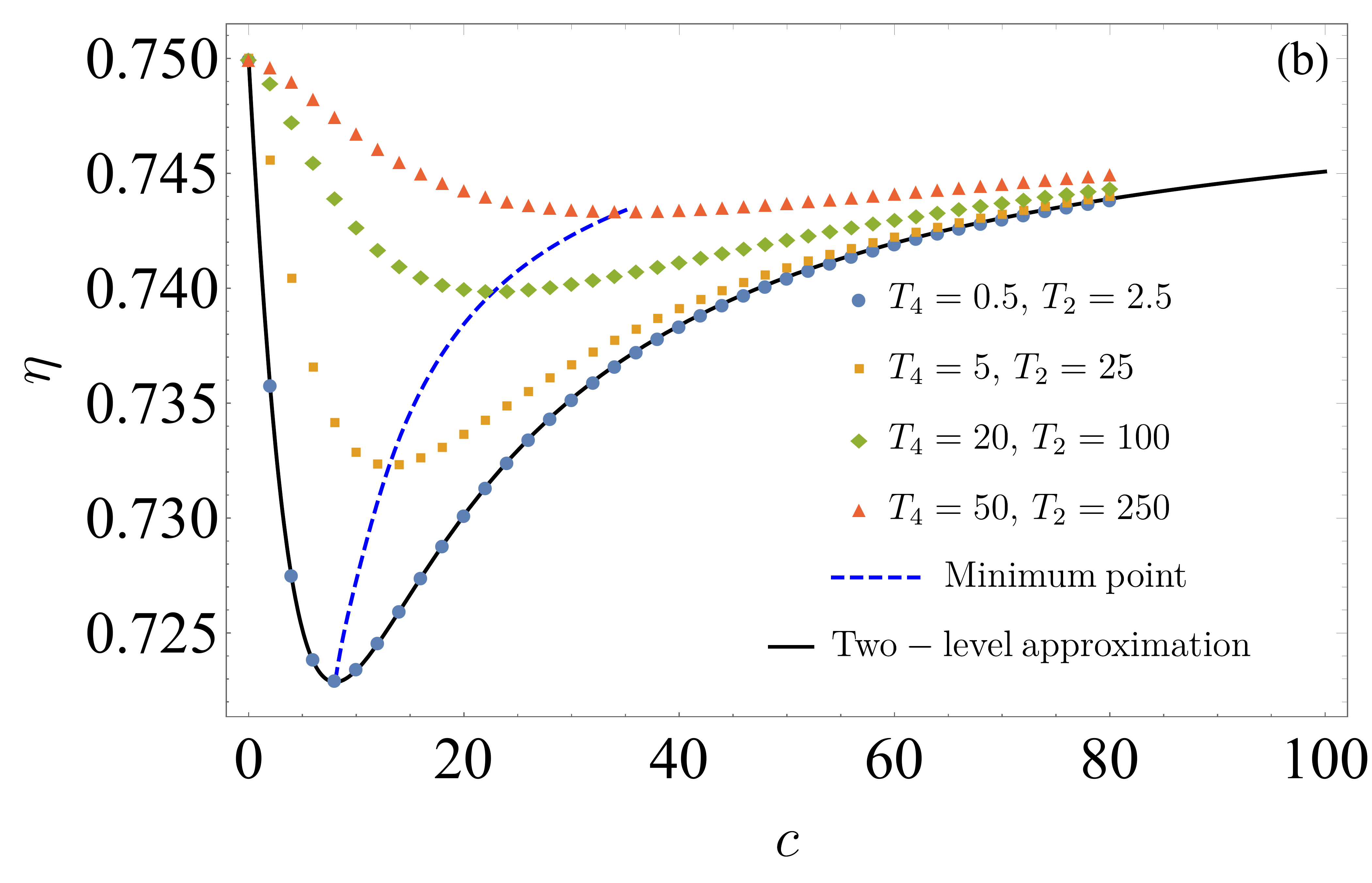}

\caption{(Color online) (a) the ratio $\lambda_{\{n_{i}\}}(L_{1},L_{2})=1-E_{\{n_{i}\}}^{(L_{2})}/E_{\{n_{i}\}}^{(L_{1})}$
with different interacting strength $c$ for two interacting Bosons
$N=2$. We calculate the ratio for the state with the quantum number
$\left(n_{1},n_{2}\right)=\left(1,1\right),\left(1,2\right),\left(1,3\right),\left(2,2\right),\left(2,3\right),\left(3,3\right).$
The solid lines are the ratios for 6 corresponding energy levels,
while the dashed line is the Carnot efficiency $\eta_{C}=1-L_{1}^{2}/L_{2}^{2}=0.75$.
The red dots are the minimum point of the ratio, not only for the
plotted energy levels but also including energy levels with higher
energy. Fig.\ref{5energyratio} (b) shows the efficiency derived by
numerical calculation for different temperature and two-state approximation
result. They match well under low temperature limit $T_{4}=0.5,\,T_{2}=2.5$.
The dots of different colors is the exact numerical result for different
temperature, while the black solid line is derived by the two-level
approximation from Eq. (\ref{eq:10twolevel}).}

\label{5energyratio}
\end{figure}

Interestingly, the curve for efficiency shows a dip with particular
interacting strength $c$ in Fig. \ref{2_loglinearplottwoparticles}.
To understand the appearance of such dip, we rewrite the efficiency
in Eq. (\ref{eq:6}) as 
\begin{equation}
\eta=\frac{\sum_{\{n_{i}\}}\left(p_{\{n_{i}\}}^{(2)}-p_{\{n_{i}\}}^{(4)}\right)E_{\{n_{i}\}}^{(L_{1})}\lambda_{\{n_{i}\}}(L_{1},L_{2})}{\sum_{\{n_{i}\}}\left(p_{\{n_{i}\}}^{(2)}-p_{\{n_{i}\}}^{(4)}\right)E_{\{n_{i}\}}^{(L_{1})}},
\end{equation}
where $\lambda_{\{n_{i}\}}(L_{1},L_{2})=1-E_{\{n_{i}\}}^{(L_{2})}/E_{\{n_{i}\}}^{(L_{1})}$
is a ratio, similar to the Otto efficiency for a two-level heat engine\cite{Quan2007}.
In Fig. \ref{5energyratio}(a), we plot the ratio $\lambda_{\{n_{i}\}}(L_{1},L_{2})$
as a function of the interacting strength for different energy levels
$\{n_{i}\}$. The curves for different energy levels show dips with
different positions. For low temperature, since the particle occupation
on higher energy levels can be neglected, we use two-level approximation
to calculate the efficiency
\begin{equation}
\eta=1-\frac{\Delta_{2}}{\Delta_{1}},\label{eq:10twolevel}
\end{equation}
where only the ground state and the first excited state are considered
with the energy gap $\Delta_{i}=E_{(2,1)}^{(L_{i})}-E_{(1,1)}^{(L_{i})},\,i=1,2$.
Under the low temperature limit ($T_{2}=2.5,\,T_{4}=0.5$ in Fig \ref{5energyratio}
(b) ), the efficiency derived with two-level approximation (black
solid curve) matches with the exact numerical result (blue dots).
For high temperature, the efficiency contains more contribution from
high energy levels, and the efficiency approaches the ideal case $1-L_{1}^{2}/L_{2}^{2}$. 

\begin{figure}[h]
\includegraphics[scale=0.2]{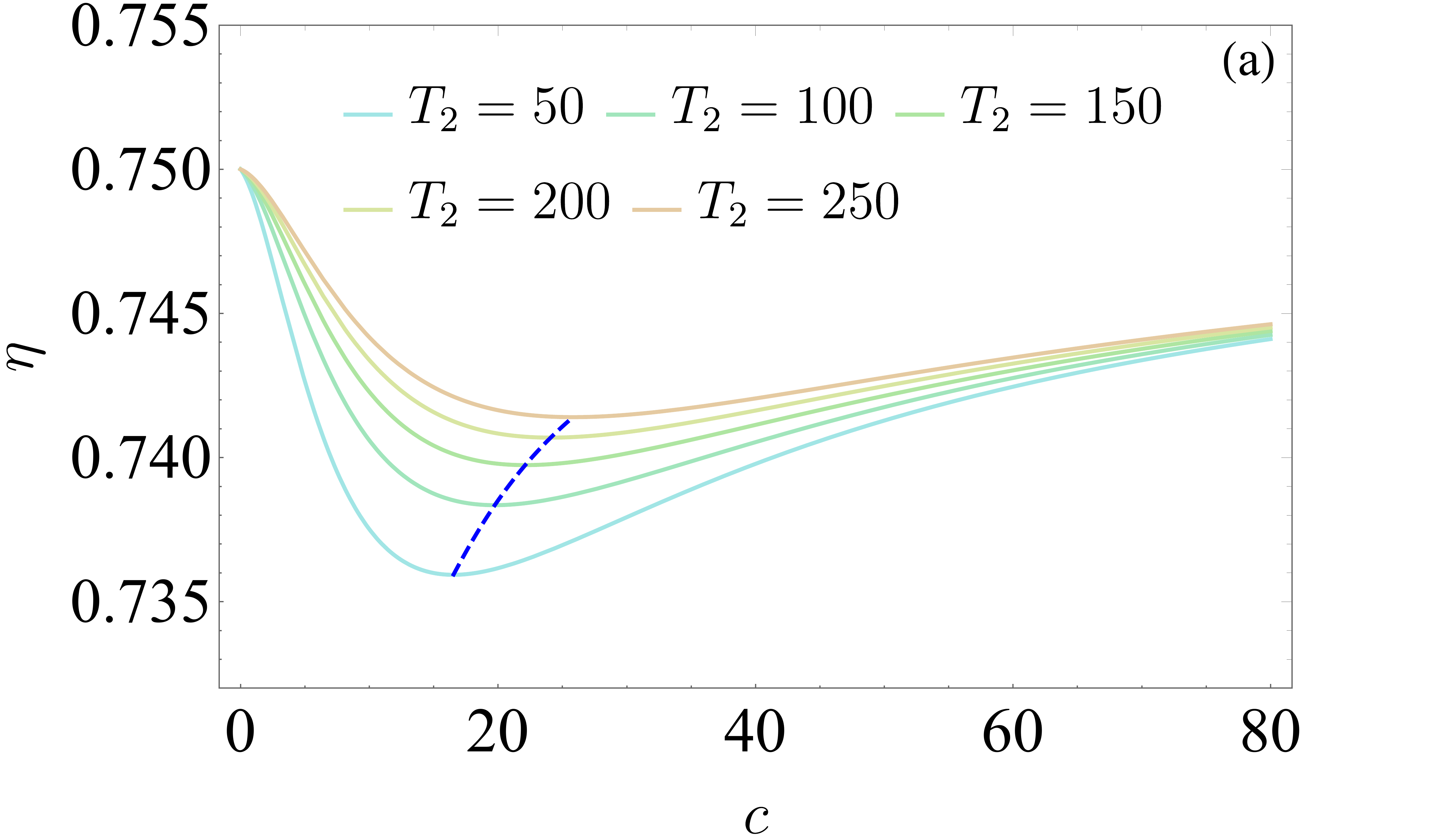}

\includegraphics[scale=0.2]{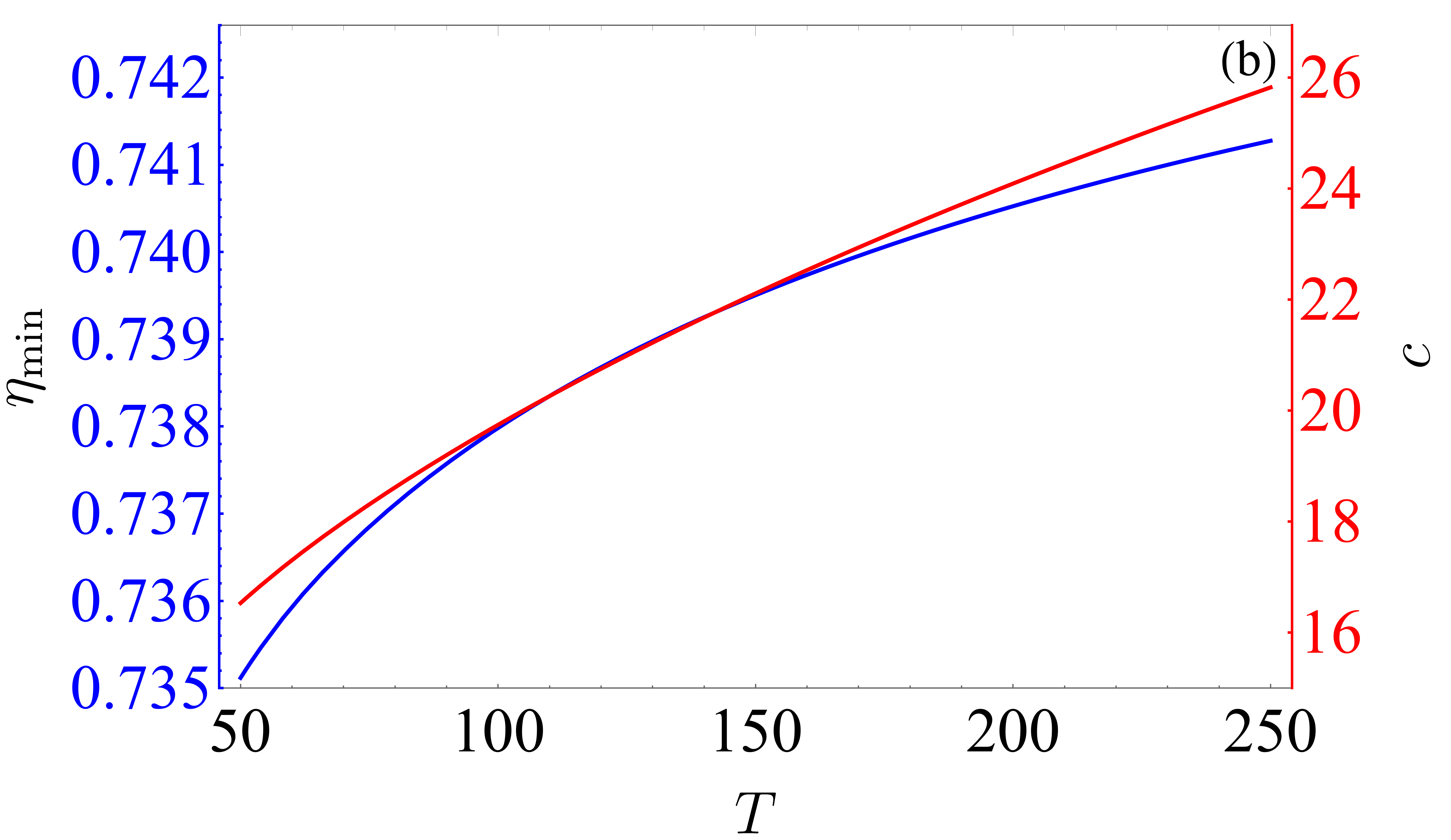}\caption{(Color online) The effect of temperature on the efficiency dip. The
efficiency-interaction curve with different temperature of the hot
reservoir $T_{2}$, modulating from $50$ to $250$. The temperature
of the cold reservoir is fixed $T_{4}=8$. In Figure \ref{fig4} (a),
the solid lines of different colors represent the efficiency $\eta$
for different temperature $T_{2}$. The blue dashed line shows the
the minimum point of the instant efficiency for different temperature
$T_{2}$. Figure \ref{fig4} (b) extracts the coordinate of the minimum
point for different $T_{2}$ in Fig. \ref{fig4}. The blue line and
the red line give the minimum efficiency $\eta_{\mathrm{min}}$ and
the interacting strength $c$ for the the minimum point for different
$T_{2}$ respectively.}

\label{fig4}
\end{figure}

The efficiency of this Otto heat engine is affected by the temperature
of the reservoirs, which is different from one-particle Otto heat
engine. In Fig. (\ref{fig4}), we study the temperature effect by
modulating the temperature $T_{2}$ of the hot reservoir from $50$
to $250$ with fixed temperature of the cold reservoir at $T_{4}=8$.
Figure \ref{fig4} (a) shows the efficiency $\eta$ is larger for
higher temperature $T_{2}$ as expected. The minimum point of the
efficiency for different temperature $T_{2}$ is plotted with red
dashed line. To figure out how the temperature $T_{2}$ affects the
minimum point of the efficiency, we plot both the efficiency $\eta_{\mathrm{min}}$
and the interacting strength $c$ of the minimum point with different
temperature $T_{2}$ in Figure \ref{fig4} (b). The interacting strength
$c$ (the red line) and the efficiency $\eta_{\mathrm{min}}$ (the
blue line) for the minimum efficiency become larger when $T_{2}$
increases, which matches the minimum point of $\lambda_{\{n_{i}\}}(L_{1},L_{2})$
in Fig. \ref{5energyratio}. The behavior of the efficiency with different
temperature $T_{2}$ matches with the change of the ratio $\lambda_{\{n_{i}\}}(L_{1},L_{2})$
of the corresponding energy levels.

\section{Conclusion}

We have studied the quantum Otto heat engine with 1D repulsive Bose
gas in a hard wall box to reveal the effect of interaction on the
efficiency. For weak interaction, we conclude that the efficiency
of the Otto heat engine is lower than the non-interacting case, smaller
for high temperature and larger for low temperature. When the interaction
is large, the efficiency recovers to its initial value, which is explained
by the Bose-Fermi duality for 1D interacting Bose gas. By calculating
the ratio $\lambda_{\{n_{i}\}}(L_{1},L_{2})$, we have explained the
appearance of the minimum value of efficiency as the function of the
interacting strength $c$. For the low temperature case, the Otto
engine with the lowest two levels gives a good approximation with
different interacting strength. For the high temperature case, the
contribution for high levels shifts the minimum position as well as
the corresponding efficiency.  

\bibliographystyle{apsrev4-1}
\bibliography{heatengine,Otto_Interaction_add}

\end{document}